\documentclass[preprint,journal]{vgtc}            

\usepackage{listings}
\usepackage{xspace}
\usepackage{amsmath,amsfonts}
\usepackage{array}
\usepackage{float}
\usepackage{tabularx}
\usepackage{amssymb}

\AtBeginDocument
{
    \definecolor{researcher_color}{HTML}{FF735E}
    \definecolor{internal_color}{HTML}{5581E0}
    \definecolor{kim_green}{HTML}{4CBB17}
    \definecolor{z_pink}{HTML}{FF69B4}
    \definecolor{sbb_purple}{HTML}{9300FF}
    \definecolor{ari_red}{HTML}{EC1313}
    \definecolor{klaus_blue}{HTML}{099AE7}
    \newcommand{\researcher}{\textcolor{researcher_color}}
    \newcommand{\internal}{\textcolor{internal_color}}
    \newcommand{\para}[1]{\vspace{0.35em}\noindent\normalsize\textbf{#1.}\xspace}

    \newcommand{\fullsys}[0]{Explainable XR\xspace}
    \newcommand{\sys}[0]{EXR\xspace}
    \newcommand{\supple}[0]{Supplementary Materials\xspace}
    \newcommand{\evalsubjects}[0]{participants\xspace}
    \newcommand{\capitalevalsubjects}[0]{Participants\xspace}
    
    \newcommand{\halfcirclesymbol}{\protect
    \tikz[baseline=-0.6ex] {
        \fill[black] (0,0) -- (0,2.5pt) arc[start angle=90, end angle=-90, radius=2.5pt] -- cycle;
    }}
    \newcommand{\circlesymbol}{\protect\tikz \fill[black] (0,0) circle (2.5pt);}
}
\onlineid{1393}

\ieeedoi{10.1109/TVCG.2025.3549537}

\vgtccategory{System}

\vgtcpapertype{System}

\title{Explainable XR: Understanding User Behaviors of XR Environments using LLM-assisted Analytics Framework}

\author{%
  \authororcid{Yoonsang Kim}{0009-0006-2341-3862},
  \authororcid{Zainab Aamir}{0009-0006-2000-6823}, 
  \authororcid{Mithilesh Singh}{0009-0007-6477-1495}, 
  \authororcid{Saeed Boorboor}{0000-0001-6644-5983}, 
  \authororcid{Klaus Mueller}{0000-0002-0996-8590}, and \\ 
  \authororcid{Arie E. Kaufman}{0000-0002-0796-6196}, \textit{Fellow, IEEE}
}

\authorfooter{
  All authors are with Center for Visual Computing at Stony Brook University, New York. 
  E-mail: \{yoonsakim, zaamir, mkssingh, sboorboor, mueller, ari\}@cs.stonybrook.edu.
}

\abstract{%
  We present \textit{\fullsys}, an end-to-end framework for analyzing user behavior in diverse eXtended Reality (XR) environments by leveraging Large Language Models (LLMs) for data interpretation assistance. Existing XR user analytics frameworks face challenges in handling cross-virtuality -- AR, VR, MR -- transitions, multi-user collaborative application scenarios, and the complexity of multimodal data. \fullsys addresses these challenges by providing a virtuality-agnostic solution for the collection, analysis, and visualization of immersive sessions. We propose three main components in our framework: (1) A novel user data recording schema, called User Action Descriptor (UAD), that can capture the users' multimodal actions, along with their intents and the contexts; (2) a platform-agnostic XR session recorder, and (3) a visual analytics interface that offers LLM-assisted insights tailored to the analysts' perspectives, facilitating the exploration and analysis of the recorded XR session data. We demonstrate the versatility of \fullsys by demonstrating five use-case scenarios, in both individual and collaborative XR applications across virtualities. Our technical evaluation and user studies show that \fullsys provides a highly usable analytics solution for understanding user actions and delivering multifaceted, actionable insights into user behaviors in immersive environments. 
}


\teaser{
  \centering
  \vspace{-1mm}
  \includegraphics[width=\linewidth]{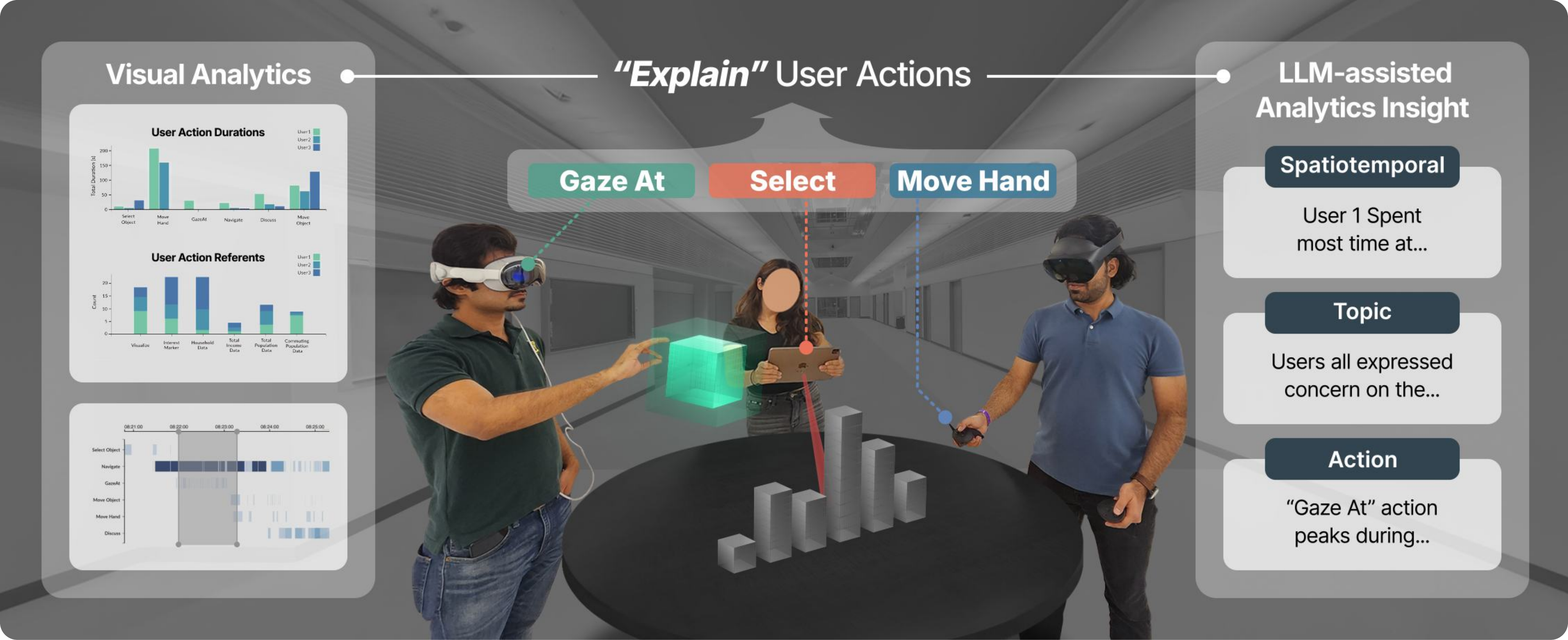}
  \vspace{-6.3mm}
  \caption{Explainable XR provides a streamlined pipeline to record, visualize, and analyze users of an immersive session, facilitating researchers from various domains of expertise, to readily comprehend and study them. Our structured user data recording format captures users' actions (e.g., GazeAt, Move Hand, Select) and their contextual reasoning. Through our data analyzer and LLM-generated insights, we present base analytics interpretation of the users' action data on top of our visual analytics interface, and assist researchers to approach the analysis from multifaceted perspectives.}
  \vspace{-2mm}
  \label{fig:teaser}
}

\graphicspath{{figs/}{figures/}{pictures/}{images/}{./}} 

\usepackage{mathptmx}                  

\begin{document}


\firstsection{Introduction}

\maketitle


The recent advancements in eXtended Reality (XR) technologies -- Augmented Reality (AR), Virtual Reality (VR), and Mixed Reality (MR) -- have significantly enhanced user experiences. 
It has transformed the way we retrieve information, interact with data, and collaborate with peer users.
As a result, immersive technologies and their applications are being widely adopted in various multidisciplinary domains such as education, healthcare, and industry~\cite{boorboor2023voxar, boorboor2023submerse, mirhosseini2019immersive, gasques2021artemis, jin2022will, yu2022duplicated, mirhosseini2019exploration}. Compared to traditional interaction methods, immersive interaction techniques have uncovered diverse and complex patterns of user behaviors and responses, highlighting the unique ways users engage with each other and the environment ~\cite{khurana2024just, mirhosseini2019immersive, saffo2021remote, wolf2018performance, sun2018towards, hu2019reducing, hu2022spatial}. Recent studies on analyzing user actions within XR environments~\cite{castelo2023argus,  hubenschmid2022relive, javerliat2024plume, nebeling2020mrat} have made significant contributions in capturing, visualizing, and interpreting user behavior data.

\vspace{-1mm}
In addition to enabling rich, immersive experiences, XR technologies also offer the capability to transition seamlessly between different virtualities within a single application. This flexibility has spurred a growing interest amongst researchers, leading to a number of studies on cross-virtuality experiences and multi-user collaboration~\cite{kasahara2012second, piumsomboon2017covar, tian2023using, lee2020shared, wang2024collabxr, enriquez2024evaluating}. As user roles, interaction patterns, and behaviors in XR environments become increasingly complex and diverse, there is a pressing need for a unified standard that accommodates various types of immersive experiences and enables consistent, systematic evaluation, analytics, and visualization across them. Moreover, a single session of an XR application can generate a large volume of multimodal data, including spatial, temporal, visual, and audio. As more sessions are integrated, the volume and complexity of the data increases, posing additional challenges in deriving insightful visualizations and interpretations due to data overload. 

To address these challenges, we present \textbf{\textit{Explainable XR}} (in short, \textit{\sys}), an end-to-end user behavior analytics framework for the collection, analysis, and visualization of XR user(s) in a spectrum of XR environments (\cref{fig:teaser}). Its main features are (i) User Action Descriptor (UAD) - a user action-centric standardized structured schema for data recording; (ii) an easy-to-use, Unity-based platform-agnostic XR session recorder; (iii)  a web-based visual analytics interface presenting multimodal data -- spatial, temporal, visual, and audio -- of user interactions of XR sessions in a single view; and (iv) leveraging Large Language Models (LLMs) to facilitate data analysis. We list our main contributions as:
\begin{itemize}
\item A publicly available~\textbf{\footnotemark[1]}, end-to-end XR user action analytics framework that supports a wide range of XR environments (AR, VR, MR), including cross-virtuality applications, multi-user scenarios, and various immersive platforms.
\vspace{-1.25mm}
\item A standardized user action-centric data schema that facilitates ease of use and scalability while capturing detailed information about user actions and the context surrounding each action.
\vspace{-1.25mm}
\item A Unity-based plugin that allows for easy customization and adaptation to specific user requirements.
\vspace{-1.25mm}
\item Leveraging Large Language Models (LLMs) to synthesize multimodal data and generate user-input tailored summaries and analytical insights, facilitating XR session analysis and interpretation. 
\vspace{-4mm}
\item A web-based analytics interface for visualizing the recorded data in a unified view, combined with generated insights for a more comprehensive analysis.
\vspace{-1.25mm}
\item The demonstration of versatility and usefulness of Explainable XR with five diverse XR scenarios ranging from single/multi-user, synchronous/asynchronous, to individualistic/collaborative applications.
\vspace{-1.25mm}
\end{itemize}
\vspace{-1mm}
\footnotetext[1]{GitHub Page: \url{https://github.com/yoonsang0910/ExplainableXR}}

\section{Related work}
\label{sec:related_work}
In this section, we review relevant works and highlight the unique contributions of \sys in comparison to them. \cref{tab:related_system_comparison} supplements this discussion by illustrating the gaps.

\begin{figure*}[!ht]
    \centering
    \includegraphics[width=\linewidth]{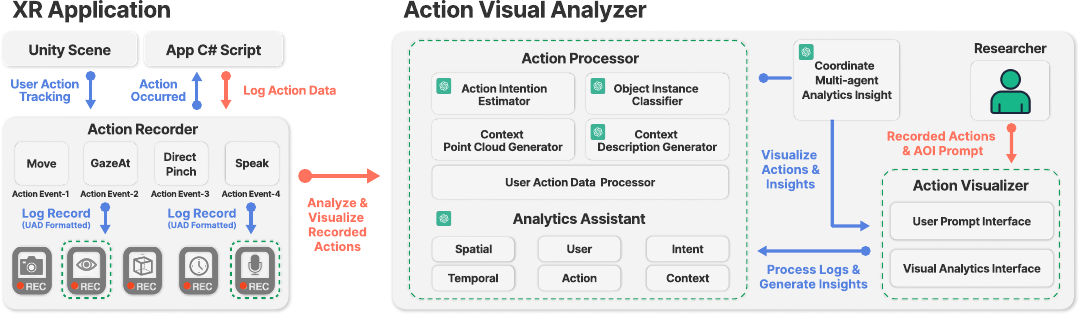}
    \vspace{-4.25mm}
    \caption{\fullsys Pipeline Overview: \textbf{\internal{The blue arrows}} denote the internal calls and flows of \fullsys, and \textbf{\researcher{red arrows}} denote the inputs of the researcher in our framework. The pipeline initiates by recording the multimodal interactions of the subjects in XR sessions, and importing it into our Action Visual Analyzer via User Prompt Interface. The researcher can perform analytical tasks in our Visual Analytics Interface and optionally utilize our analytics insights during the analysis.}
    \vspace{-2.75mm}
    \label{fig:explainable_xr_overview}
\vspace{-2.5mm}
\end{figure*}

\paragraph{\textbf{Recording of XR User Behavior Data}}
\label{related_work:user_data_recording}
Capturing various data streams in an XR session, including 6DoF (Degrees-of-Freedom) transformations, actions, gestures, and physiological data such as gaze can provide valuable insights into understanding user intentions and behavior patterns \cite{david2021towards, vivik23identify, numan2022exploring, robert2023integrated,   romero2023gaitguard, slocum2023going, zhang2023s}. The recording of such data provides the means for a deeper analysis of users and the context behind every XR interaction. 

Several toolkits and frameworks for immersive sessions have been developed to derive insights from these various user data streams. One of the early toolkits to capture user actions in AR and MR is MRAT~\cite{nebeling2020mrat}. MRAT offers user information logging per task-basis. Given a set of tasks for user performance testing, it tracks the user's task status, spatiotemporal data, interaction type, target virtual object, gaze, screenshot, gestures, and voice commands, which are then used to visualize and analyze the task performance of a user. UXF~\cite{brookes2020studying} proposes a Unity-based framework that simplifies VR experiment development and data collection for behavioral research. Other works also place emphasis on human behavioral experiments in Virtual Reality~\cite{brandstatter2023dialogues, gorisse2022rec, villenave2022xrecho}. Frameworks such as Cognitive3D~\cite{cognitive3D} advance a step further by providing support for cross-virtuality (AR, VR, MR) recording and playback. Furthermore, PLUME~\cite{javerliat2024plume} and other recent studies~\cite{deuchler2023streamlining} incorporate the ability to record users' physiological signals such as eye tracking and heart rates as well. A set of research concentrates on collecting data from a first-person perspective~\cite{cools2023crest, martinez2024clovr, wen2022vrhook}. Some works focus specifically on the multi-user aspect of XR, recording multiple users' actions~\cite{buschel2021miria, nebeling2020mrat, steed2022ubiq}.

The above works and \sys record similar XR multimodal data such as visual, gaze, and audio, as well as user actions, context, spatial and temporal data. Our work, however, takes a more structured, user action-centric approach. We consider user actions as a trigger for an information update in an XR environment, and we link all tracked XR data to a user's action. This logging structure allows our \sys to capture user behavior and relevant context, while filtering out non-essential session data. Additionally, this user action-centric approach allows for scalability in multi-user scenarios.
\vspace{-0.5mm}

{
\renewcommand{\arraystretch}{1.2}
\setlength{\tabcolsep}{2pt}
\begin{table}[t]
\normalsize
\centering
\caption{
Comparison of XR user analytics systems. Symbols (\halfcirclesymbol, \circlesymbol) denote partial and full functionality support of a system, respectively. Action Context refers to the built-in capability to save the context behind users' actions. Cross-user indicates support for single and multi-user sessions. Task-agnosticism highlights the systems adaptability to diverse tasks, rather than being tailored to a specific task. Novel functionalities exclusive to \sys are omitted.
}
\vspace{-3mm}
\label{tab:related_system_comparison}
\begin{tabular}{| l | l | c | c | c |}
\hline
\textbf{System} & \textbf{Virtuality} & \textbf{Action} & \textbf{Cross-} & \textbf{Task-} \\
 &  & \textbf{Context} & \textbf{user} & \textbf{agnostic} \\

\hline
ARGUS~\cite{castelo2023argus} & AR & \tikz \fill[black] (0,0) circle (2.5pt); &  &  \\

PLUME~\cite{javerliat2024plume} & AR, MR, VR & & \tikz[baseline=-0.6ex] {
    \fill[black] (0,0) -- (0,2.5pt) arc[start angle=90, end angle=-90, radius=2.5pt] -- cycle;
} & 
\tikz[baseline=-0.6ex] {
    \fill[black] (0,0) -- (0,2.5pt) arc[start angle=90, end angle=-90, radius=2.5pt] -- cycle;
} \\

MIRIA~\cite{buschel2021miria} & AR, MR &   & \tikz[baseline=-0.6ex] \fill[black] (0,0) circle (2.5pt); &  
\tikz[baseline=-0.6ex] {
    \fill[black] (0,0) -- (0,2.5pt) arc[start angle=90, end angle=-90, radius=2.5pt] -- cycle;
} \\

MRAT~\cite{nebeling2020mrat} & AR, MR & \tikz[baseline=-0.6ex] {
    \fill[black] (0,0) -- (0,2.5pt) arc[start angle=90, end angle=-90, radius=2.5pt] -- cycle;
} & \tikz \fill[black] (0,0) circle (2.5pt); & 
\tikz[baseline=-0.6ex] {
    \fill[black] (0,0) -- (0,2.5pt) arc[start angle=90, end angle=-90, radius=2.5pt] -- cycle;
} \\

ReLive~\cite{hubenschmid2022relive} & AR, MR, VR & \tikz[baseline=-0.6ex] {
    \fill[black] (0,0) -- (0,2.5pt) arc[start angle=90, end angle=-90, radius=2.5pt] -- cycle;
} & \tikz \fill[black] (0,0) circle (2.5pt); &  
\tikz[baseline=-0.6ex] {
    \fill[black] (0,0) -- (0,2.5pt) arc[start angle=90, end angle=-90, radius=2.5pt] -- cycle;
} \\

\textbf{EXR (Ours)}& AR, MR, VR & \tikz \fill[black] (0,0) circle (2.5pt); & \tikz \fill[black] (0,0) circle (2.5pt); & \tikz \fill[black] (0,0) circle (2.5pt); \\
\hline
\end{tabular}
\vspace{-5.75mm}
\end{table}
}

\paragraph{\textbf{Visualization of XR User Behavior Data}}
\label{related_work:user_data_vis}
Effective visualization of an XR session is crucial for visual interpretation and the comprehension of user behaviors. There have been two main approaches: In-situ and ex-situ. The former, often used with immersive analytics, focuses on first-person views to trace users' reasoning processes~\cite{buschel2021miria, luo2023pearl, nebeling2020mrat}. Ex-situ visualizations are typically paired with 2D visual analytics interfaces, and use third-person views to provide a broader understanding of users' actions and the contexts behind their actions~\cite{brudy2018eagleview, cognitive3D, hubenschmid2022relive}. Some studies take advantage of both and combine immersive in-situ and non-immersive ex-situ analysis~\cite{jansen2023autovis, javerliat2024plume}.

ARGUS from Castelo et al.~\cite{castelo2023argus} provides an analytics interface for visualizing AI model outputs of an AR session. They offer both real-time (online) and retrospective (offline) tracking of the AR user's scene and interactions, along with debugging functionalities. The work focuses on the scenario where a user with an AR Head-mounted Display (HMD) performs physical tasks through an AI-guided system, analyzing spatiotemporal properties of user actions as well as the AI model's outputs. Also, to visualize the physical environment (action context) at the moment of a user action, they use sparse point clouds on top of screenshot captures, similar to Yu et al.~\cite{yu2023dynascape}. ReLive~\cite{hubenschmid2022relive} proposes a visual analytics interface to provide a holistic visualization of individual user actions and a synchronized, aggregated view of multiple users. Their data logging toolkit captures and stores the action context of users -- 3D scene for virtual applications (VR) and accept real-world scan for physical applications (AR, MR). Also, they maintain screenshots for action events.

Existing systems often use visual elements such as graphs, plots, glyphs, or attention maps (2D or 3D) to enhance the understanding of user's actions by visualizing observable data. \sys extends this approach by not only tracking and visualizing user behaviors but also providing insights into the potential reasoning behind each action, powered by LLM. This provides analysts with a deeper understanding into the motivations and contexts behind XR user interactions. By integrating spatial, temporal, and interaction data, \sys allows for a comprehensive analysis of XR session dynamics. 

\paragraph{\textbf{AI-assisted Analytics and Visualization}}
\label{related_work:ai_analytics}
The integration of Artificial Intelligence (AI) through Large Language Models (LLMs) has opened new avenues for data analysis and visualization. This has enabled advanced pattern recognition, predictive analytics, automated insight generation, and context-aware visualization, enhancing the data analytics experience~\cite{bohus2024sigma, qu2024llms, shen2022towards, wu2021ai4vis}.

Recent studies have explored various AI-assisted methods to enhance data analytics.  InsightPilot~\cite{ma2023insightpilot} proposes a streamlined data exploration process that generates  data insights reducing the effort needed to understand the data~\cite{choe2024enhancing, nam2024using,  zhao2024leva}. LIDA~\cite{dibia2023lida} proposes an infographics generation pipeline using LLM to suggest visualizations. Other works have also utilized AI for generating visualizations~\cite{text2viz, de2024llmr, dogan2024augmented}. While Shen et al.~\cite{shen2024data} applies task decomposition with multiple LLM agents for automatic data storytelling generation. Although AI-powered data analysis and visualization offers powerful capabilities, it present challenges on reliability and trustworthiness of the outputs, leading to research on the transparency and explanability of AI generated output~\cite{lubos2024llm}.

Building on these developments, \sys incorporates AI/LLM-assisted analytics optimized specifically to our user action-centric data structure. We maintain the human-in-the-loop approach to ensure reliable data exploration and decision making. In \sys, LLM is used to understand large, multivariate datasets, emphasizing key information and patterns. The use of LLM is underscored as \sys manages complex visual analytics involving multiple users across diverse XR configurations and virtualities.

\paragraph{\textbf{Task-agnostic User Analytics Framework for XR}}
\label{related_work:multi_purpose_framework}
Numerous studies have focused on designing analytical frameworks for XR sessions~\cite{brudy2018eagleview, castelo2023argus,  cognitive3D, hubenschmid2022relive, javerliat2024plume, nebeling2020mrat}. These works commonly provide XR session recorder and a viewer, but most are tailored for specific tasks such as measuring user performance or testing ~\cite{brookes2020studying, hubenschmid2022relive, nebeling2020mrat}, or an AI-guidance system~\cite{castelo2023argus, castelo2024hubar}, although they can be customized to a degree. PLUME~\cite{javerliat2024plume} proposes a more open-ended framework that is not bound to a specific task. It utilizes low-level compile time code modifications to track log event raises for logging XR session data. Additionally, PLUME provides both in-situ and ex-situ visualization of the XR session. However, it lacks scalability for multi-user scenarios and practical AR/MR applications involving physical scenes with no prior context information (pre-scanned scene).

As compared to other works, we extend the use-case of our framework beyond user performance measurement by introducing a versatile action-centric data structure. This supports diverse tasks such as spatial, temporal, topic, and action intention analyses, positioning \sys as a general-purpose analytics framework for XR sessions.

{
\renewcommand{\arraystretch}{1.2}
\begin{table*}[!ht]
\normalsize
\centering
\caption{User Action Descriptor: This structure organizes all immersive session data centered around a subject's actions, enabling \sys to analyze multivariate connections between actions and their surrounding contexts. It is adaptable for any immersive application or task across virtualities.}
\label{tab:action_descriptor}
\vspace{-3mm}
\begin{tabular}{|l|l|l|l|}
\hline
\textbf{\rule{-2pt}{2.5ex} Field} & \textbf{\rule{-2pt}{2.5ex} Description} & \textbf{\rule{-2pt}{2.5ex} Data Type} & \textbf{\rule{-2pt}{2.5ex} Example Value} \\
\hline
Name & The name of the action & String & ``Navigate'', ``GazeAt'', ``Touch'' \\
Type & The type of the action & Enum (Type of Action) & Discrete, Continuous \\
Intent & The intent behind the action & String & ``Load immersive plots'' \\
User & The user identity of the action invocation & String & ``User1'' \\
Location & The 6DoF locations of the action invocation & List${<}$Transform${>}$ & [(Pos(0,0,0), Rot(0,5,5)),..] \\
TriggerSource & The medium on which the action is triggered & Enum (InputAction Device) & XRHMD, XRController, Audio \\
StartTime & The start time of the action event & TimeStamp (Ymd${:}$HMS${:}$f) & 240801${:}$092855${:}$031 \\
Duration & The lengths of the action event & TimeDelta (Ymd${:}$HMS${:}$f) & 000000${:}000135${:}328 \\
Referent & The target object of the action & Bytes (GLB or PNG) & GameObject.glb, Screenshot.png \\
ReferentType & The reality in which the target object exists & Enum (Type of Reality) & Physical, Virtual \\
ReferentLocation & The 6DoF locations of the target object & List${<}$Transform${>}$ & [(Pos(10,5,4), Rot(0,-5,5)),..] \\
Context & The context behind the action & Bytes (GLB) & PointCloud.glb \\
ContextType & The reality in which the context exists & Enum (Type of Reality) & Physical, Virtual\\
\hline
\end{tabular}
\vspace{-5mm}
\end{table*}
}

\vspace{-1mm}
\section{Explainable XR: Design and Implementation}
\label{sec:exp_xr}
Analyzing behavioral patterns is critical for understanding users and task efficiency in XR environments. 
\sys is designed to facilitate domain experts from diverse backgrounds interested in studying human subjects in XR settings, to be able to readily collect, visualize, and analyze the behavioral patterns of the subjects of the immersive sessions. As an end-to-end framework, it provides base template designs for tasks spanning from recording to analysis, with the ability to customize any functionality. It is designed on top of Unity3D~\cite{unity}, a widely used engine for XR application development.
\cref{fig:explainable_xr_overview} illustrates the pipeline of our framework. In designing our framework, we abstract the convoluted inner workings of data processing and the LLM logistics.

As illustrated in~\cref{tab:related_system_comparison}, \sys is designed as an ``all-in-one package'' framework that bridges the gap between existing XR user analytics systems. It is general-purpose (not confined to a specific task), operates seamlessly across virtualities, supports both single and multi-user sessions, inherently embeds contextual information (environment) with every user action, and offers intelligent assistive techniques to enhance the analytics experience.

The process begins with the \hyperref[subsec:recorder]{Action Recorder}, which can log the actions of the users such as `Move,' `Grab Object,' or `GazeAt.' 
Subsequently, each log is grouped by user and action, and structured using our User Action Descriptor (UAD) schema. This descriptor can be used across XR virtualities and configurations. Next, the structured data is loaded using a User Prompt Interface, along with an optional Analysis-of-Interest (AoI) prompt. Providing an AoI allows \sys to generate user-interest tailored insights using an LLM. Then, through our \hyperref[fig:analytics_interface]{Visual Analytics Interface (VAI)}, they can visually analyze user task performance and patterns. The components of VAI are interconnected, allowing for a unified analysis where selections from one component update the others. Henceforth, in this paper, we interchangeably use the terms user behavior, action, and interaction of XR. Moreover, we represent the user using the XR application and being recorded, as a \textit{\textbf{Subject}}, and the user of \sys analytics framework as \textit{\textbf{Analyst}}. The term user in naming the \sys components refers to the subjects.

\vspace{-0.5mm}
\subsection{User Action Descriptor}
\label{subsec:action_descriptor}
UAD is an action-centric structure that preserves various aspects of a subject interacting with an individual or collaborative XR application session.
As discussed in \cref{sec:related_work}, existing works loosely define event-trigger conditions for data logging. Any occurrence of an event in the session, including events that are irrelevant to the subject's view or action in the XR session space, can be recorded. For example, a moving cube in a VR session can be tracked and stored throughout, even when it is out of the user's viewing frustum. This can result in data overload, for both processing and analysis. The information presented in XR applications is typically egocentric. That is to say, the presented visualizations, flow of information, and interactions initiate from a subject's point-of-view (PoV) when the subject performs an action with the XR environment.
This is also true for third-person AR applications, as they rely on the user-induced in-application camera (virtual camera) position and orientation. Let's consider a practical AR application scenario:

\vspace{1.25mm}
\noindent
``\textit{Jake entered the kitchen wearing an AR HMD to cook a dish with the help of an AI-guided AR application.}''
\vspace{1.25mm}

Loading the recipe data was triggered by Jake's \underline{entrance} (\textbf{User Action1}) into the kitchen, and Jake \underline{selected} (\textbf{User Action2}) one of the options from the cooking recipe on the AR User Interface, which gray-highlighted the interface button. Also, Jake \underline{tapped} (\textbf{User Action3}) on the button of the interface for the selection. This indicates that a subject's action was the source of all information in XR. In other words, a subject's action triggers an update in an XR environment.

Drawing from that, we have designed the UAD to consolidate the collected session data based on the subject's actions, gestures, or behavior.
These include examples such as navigating, touching an AR-projected object, pinching a virtual cube, or gazing at another subject. The UAD inherits its base concept from the Kipling method, 5W1H -- When, Where, Who, What, Why, and How -- to describe the context and the information of an incident~\cite{usercentric5w, yu2010study, Cao20245W1HEW}. Below, we outline the association of the Kipling method (in bold) with the schema of each UAD field (in italics).
The complete list of the UAD fields is provided in \cref{tab:action_descriptor}.

 \vspace{0.4em}

\noindent\textbf{When:} analyzes the temporal property of an action. We record the occurrence of an action with \textit{StartTime} and \textit{Duration}.
 \vspace{0.45em}

\noindent\textbf{Where:} analyzes the spatial property of an action. We track the 6DoF of the invoked action and the 6DoF of the action's target, stored as \textit{Location} and \textit{ReferentLocation}, respectively.
 \vspace{0.45em}

\noindent\textbf{Who:} traces the source of an action, stored as the \textit{User}. This information is especially useful for analyzing collaboration in multi-user XR scenarios. We enforce subject anonymity and assign \textit{User} with a numerical identifier for separating unique subjects.
 \vspace{0.45em}

\noindent\textbf{What:} defines the action within the XR space, categorized by its \textit{Name} and \textit{Type}. The \textit{Type} can be discrete (`Button Press', or `Pinch') or continuous (`Move Object' or `GazeAt'). Unlike discrete, continuous actions involve a sequence of spatial movements within an action.
\vspace{0.45em}

\noindent\textbf{Why:} identifies the \textit{Intent} of an action, allowing a single action to have more than one usage in an application. For instance, a single `pinch' action can be used for ``grasping an object'' or ``initiating a teleportation'', as defined by the application developer. Moreover, \textit{Intent} describes the specific consequences of an action as well. Thus, we leverage this to track the reasoning behind the subject's action, in the later stage of \hyperref[para:action_intent_estimation]{Action Intention Analysis} and inference.
\vspace{0.45em}
 
\noindent\textbf{How:} specifies the method used to trigger an action, recorded in \textit{TriggerSource} by tracking the sensor or module recognizing the action. 
For seamless use across XR environments and platforms, we utilize Unity’s Input System~\cite{unityInputSystem}, inheriting all trackable sensors and modules such as `HandheldARInputDevice', `XRHMD', and `XRController'. 
 \vspace{0.45em}

\noindent\textbf{And More:} identifies additional visual cues of the action to provide additional context for an action. In the UAD, the \textit{Referent} denotes the interactable target of an action. We store the \textit{Referent} using GLB, a platform-agnostic format, to support the use of UAD across all XR environments. It can store the name, geometry, and material of any virtual entity in Unity. To even support the storage of a physical entity in the \textit{Referent} field, we classify the physical referent via our post-hoc processing module, \hyperref[para:action_referent_classifier]{Action Referent Classifier}, and store it.

We also maintain the circumstantial context of an action. The \textit{Context} field stores the semi-dense 3D point cloud of the subject's XR scene, at the time of an action, allowing \sys to associate each action to the interaction space/scene directly. We choose point cloud over other 3D representations, such as Neural radiance fields~\cite{mildenhall2021nerf} or Gaussian splats~\cite{kerbl20233d}, for its compatibility and portability~\cite{lee2021sharing, tian2023using}. Moreover, it is more robust than a surface mesh in maintaining the spatial information. The \textit{Context} is stored as a GLB as well and supports recording of both physical and virtual scenes. The point cloud reconstruction is done in the post-hoc processing module, \hyperref[para:context_point_cloud_gen]{Context Point Cloud Generator}. This eliminates the need to upload prior context (VR), a scanned environment (AR/MR), or a digital twin (AR/MR) of the subject's interaction space. Furthermore, since we bind the context to every action of a subject, EXR can even capture and visualize the consecutive movements of an object as long as an action was involved (e.g., Gazed At, Grabbed). To illustrate a usecase, consider the following example:

\vspace{1mm}
\noindent
``\textit{\underline{At the beginning} (\textbf{When}) of a Mixed Reality user study session, \underline{Bob} (\textbf{Who}) \underline{pressed} (\textbf{What}) a \underline{UI button} (\textbf{Referent}) that is anchored near the \underline{starting position} (\textbf{Where}), with his \underline{hand} (\textbf{How}), to \underline{visualize immersive analytics data} (\textbf{Why \& Context}).}''
\vspace{-1mm}

\subsection{User Action Recorder}
\label{subsec:recorder}
Action Recorder offers an easy-to-use plug-in for logging a subject's multimodal XR session data, simplifying the complex logistics of data recording. Optimized and tailored for the UAD format, it supports two recording methods to capture a subject's actions: Template-based logging and Direct logging.

\begin{figure}[t]
    \centering
    \includegraphics[width=\linewidth]{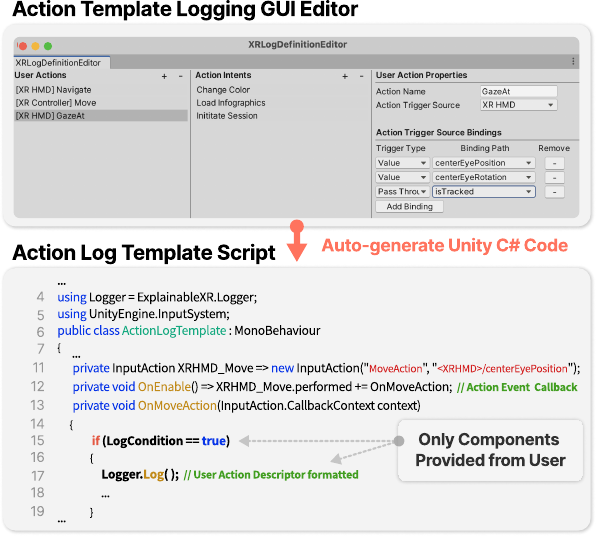}
    \vspace{-6.5mm}
    \caption{Action Template Logging Editor and its auto-generated code: Our visual editor streamlines the process of action logging by generating a Unity C\# base template code. The user can record subjects' immersive session data with a simple modification of the conditional statement (\textit{LogCondition}) and the log (\textit{Logger.Log}) function arguments. The code is generated with a button press in the visual editor, from the user.}
    \label{fig:template_based_logging}
\vspace{-1mm}
\end{figure}

\para{\textbf{Template-based Logging}}
\phantomsection
\label{para:template_logging}
To provide a starting point for developers and investigators to readily use the UAD format of \sys, we have designed a Unity editor-based GUI called Action Template Logging Editor that generates a base template C\# script for logging from input actions. As shown in \cref{fig:template_based_logging}, our visual editor can accept any number of actions and their associated intents and sources. The list of trackable devices (\textit{Trigger source}) is device-agnostic, as the recorder module can be deployed on any Unity-developed XR device. Also, the \textit{Trigger source} can be customized on top of the existing \textit{Trigger sources}. Once the developer completes the configuration of the actions on the visual editor, our framework automatically generates a script with the basic structure for action event listening, condition statements, and template codes for logging. The code generator assigns each action intent a separate function that is named after the user-specified names of the action, intent, and the \textit{Trigger source}, in the visual editor.

\para{\textbf{Direct Logging}} 
Direct Logging is a more advanced approach to recording subjects' session information. Using this approach, based on the analyst's needs, they have the freedom to insert a logging function directly into their scripts. \cref{fig:log_statement} shows an example of the logging function and its detailed arguments. This approach requires manual initialization of the Logger and data storage at the completion of a session. 

Both Template-based and Direct Logging, invoke the \textit{Log} function shown in \cref{fig:log_statement}, to record an action. Once the \textit{Log} function is invoked from the application C\# script, the defined parameters are processed to follow the UAD format. For the \textit{Referent} field, a Unity object is converted to GLB and a visual capture of a physical object is stored.
For the \textit{Context} field, the visual (RGB) and depth (Depth) images are captured using the XR device, and the camera parameters (intrinsic, extrinsic) are stored. These are subsequently used in the post-hoc processing phase to reconstruct the 3D point cloud of the subject's action context as illustrated in \cref{fig:point_cloud_generation}. Snapshot captures help minimize the overhead induced by an application in logging information such as physical referent detection, classification, audio processing, and \hyperref[para:context_desc_gen]{Action Context Image Analysis}.
\vspace{-1.1mm}

\subsection{User Action Visual Analyzer}
\label{subsec:visual_analyzer}
As illustrated in \cref{fig:explainable_xr_overview}, User Action Visual Analyzer consists of three main components: Action Processor, Analytics Assistant, and Action Visualizer. In this section, we describe how the recorded action data is processed and prepared for visual analysis, followed by an explanation of the techniques used to provide the multifaceted analyses.
\vspace{-1.1mm}

\subsubsection{Action Processor}
\label{subsubsec:processor}
The Action Processor is the core preparation stage for the visual analysis of the session data. Given the recorded data, it concatenates the action data of each subject into a single file, generates semi-dense point clouds from the context data, and performs various LLM inferences to facilitate analysis. The inference includes Action Context Description Generation, Action Intention Estimation, and Action Referent Classification. 

\begin{figure}[t]
    \centering
    \includegraphics[width=\linewidth]{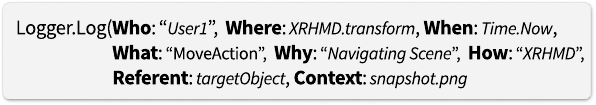}
    \vspace{-6.8mm}
    \caption{Structure of Logging Function: It conforms to the User Action Descriptor format and internally stores the action data in JSON format, upon invocation.}    
    \label{fig:log_statement}
    \vspace{-6.1mm}
\end{figure}

\begin{figure}[t]
    \centering
    \includegraphics[width=\linewidth]{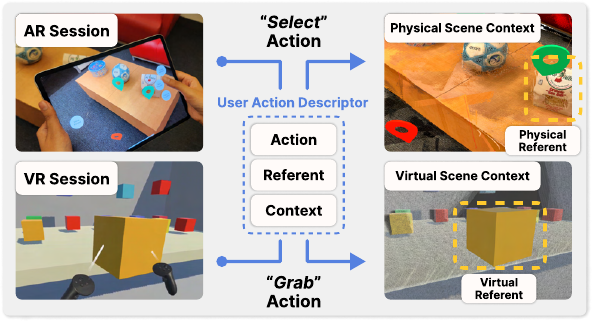}
    \vspace{-6mm}
    \caption{Virtuality-agnostic Session Reconstruction: UAD binds each action of a subject with a referent and a scene context. The referent that is a physical entity is inferred through \hyperref[para:action_referent_classifier]{Action Referent Classifier}, and the virtual entity is logged through gameobject storage. The context point cloud can be generated using the snapshots from a Unity in-application camera, or through a physical XR device camera. The former is mostly used for VR, and the latter, for AR/MR.}
    \vspace{-1mm}
    \label{fig:point_cloud_generation}
\vspace{-5mm}
\end{figure}

\vspace{-0.1mm}
\para{\textbf{Action Context Point Cloud Generation}}
\phantomsection
\label{para:context_point_cloud_gen}
Point clouds are generated to visualize the contextual depiction of the subject's surroundings at the point of an action. It provides an enhanced background reasoning of the action. As shown in \cref{fig:point_cloud_generation}, we generate a semi-dense point cloud from logged images from the logical in-application camera for VR applications and logged RGB-D images from the physical device camera for AR/MR applications. 

\para{\textbf{Action Context Description Generation}}
\phantomsection
\label{para:context_desc_gen}
To further assist data exploration and analysis, a textual description of the context information is generated. The natural language descriptions such as caption or annotation, along with the visualization, can present multimodal insights to the users~\cite{sultanum2018more, wu2021ai4vis, zhang2020viscode}. To this end, we leverage a multimodal LLM by querying the subject actions, intents, and referents, to describe a snapshot of the subject action. The textual description or annotation is then appended to the UAD \textit{Context} field.

\para{\textbf{Action Intention Estimation}}
\phantomsection
\label{para:action_intent_estimation}
As shown by the `Speak' action in \cref{fig:template_based_logging}, not all action intentions can be determined during application development. The intention of a verbal discussion (`Speak' action), can only be interpreted within the full context upon the completion of a discussion. We classify such cases as `Post Defined' intentions. To address these, we employ a multimodal LLM agent tasked with deducing the plausible intention of an action based on the subject’s context, action, visual snapshot, and transcribed verbal communication. The significance of intention deduction for verbal interactions is particularly critical in multi-user collaborative scenarios. Verbal expressions serve as explicit indicators of a subject’s reasoning~\cite{nebeling2020mrat}. When combined with other actions, they provide valuable insights into subjects’ revealed or hidden interests and behaviors.

\para{\textbf{Action Referent Classification}}
\phantomsection
\label{para:action_referent_classifier}
In AR/MR, a subject's interaction is not limited to the virtual realm as it involves augmenting information on physical objects. In practice, XR applications cannot exhaustively recognize interactable elements in the physical world. This can limit scalability and limit accuracy to VR applications. Thus, for AR/MR settings, we utilize the logged snapshot and infer the referent of the action in the post-processing by classifying the physical object using the LLM agent. The deduced object class along with its confidence score, is then added to the \textit{Referent} field.
\vspace{-1.35mm}

\subsubsection{Analytics Assistant}
\label{subsubsec:analytics_assistance}
An increased amount of presented information can negatively impact the user's ability to absorb the information due to information overload~\cite{arnold2023dealing, mahdi2020information}. To address this, \sys offers two solutions: Analytics Insight and Analysis-of-Interest Marker.

\para{\textbf{Analytics Insight}}
To facilitate efficient data exploration and analysis, a curated list of LLM-generated insights (up to 10) are presented to the analyst, which serves as entry points for analyzing extensive datasets. These insights provide a concise recapitulation of the recorded XR session, offering analysts a quick overview of key points and patterns. To customize the generated insights, we have incorporated an AoI feature, where the analyst can specify their analytical focus. The AoI is prompted at the beginning of the Action visualizer pipeline via `User Prompt Interface' shown in \cref{fig:explainable_xr_overview}.

The extracted insights cover six distinct analytical aspects: space, time, action, intent of action, context of action, and user (interaction/collaboration analysis). This multifaceted approach enables \sys to guide users through a wide range of topics and tasks, from broad queries similar to ``\textit{Insights on the discussed topics in user collaboration}'' to more specific inquiries such as ``\textit{Insights on the time spent object with Gaze action}''. 

During the insight generation process, LLM plays a critical role in synthesizing multimodal data streams such as gaze, gestures, and spatial interactions, captured in the UAD format. By contextually binding these diverse inputs, LLM generates cohesive insights that reflect the context-awareness of user behaviors in XR environments. This enhances the \textit{`explain'}ability of data analytics, enabling the identification of implicit data patterns across modalities that may be difficult to derive through unguided data exploration and analysis. The example use-cases of our LLM-assisted insights, along with AoI prompts, are illustrated in~\cref{fig:protoapps_aoi}. For details on the techniques we applied to the LLM-generated results, refer to the \supple.

\para{\textbf{AoI Marker}}
\phantomsection
\label{para:aoi_marker}
As shown in \cref{fig:analytics_insight}, the AoI Marker is an analytics guidance module that visually highlights potential interests derived from the Analytics Insight module. Each highlighted insight points to the source action from which the reasoning is derived and its timestamp marked on the \hyperref[para:temporal_viewer]{Temporal Viewer}. This visual guide enhances the user's ability to navigate and focus on pertinent data points while maintaining the original information.
\vspace{-1mm}

\subsubsection{Action Visualizer}
\label{subsubsec:analytics_interface}
The visual analytics of subject behaviors starts from importing the recorded action log to the User Prompt Interface shown in \cref{fig:explainable_xr_overview}. The primary goal of the Visual Analytics Interface is to provide multifaceted analytical views of the UAD, processed by the Analytics Insight and AoI Marker modules. 
We now describe each individual component of the Action Visualizer.

\begin{figure*}[t]
    \centering
    \includegraphics[width=\linewidth]{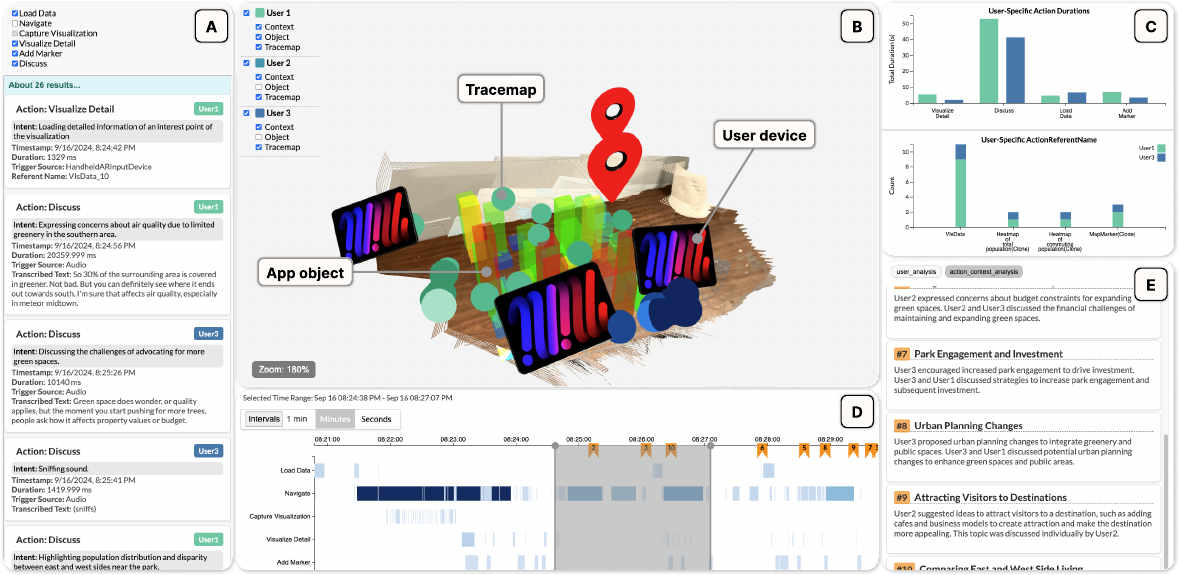}
    \vspace{-6mm}
    \caption{Visual Analytics Interface: (A) The Data Manager includes Data Filter that filters the visualized action across the viewers, and Data Viewer which provides a way to examine the raw recorded action data of the selected time range. (B) Spatial Viewer visualizes the spatial properties of the subjects' actions, referent, the context point cloud, and action traces. The Trace Map can visualize the spatial patterns of actions (e.g., trajectory, locations of collaboration). (C) Plot Viewer visualizes the statistics of the subjects' interacted objects (Referent), and the duration of each action. (D) Temporal Viewer is the main control point to the information shown across the viewers. Users can select the sampling bin size of the time steps, or the time range, and interact with the AoI Marker to locate the interest points of their AoI. (E) Insight Viewer shows the LLM-generated insights facilitating the users' analytics and reasoning. The `User device' in this example is an iPad used from \hyperref[sec:app_scenarios]{A4}.}
    \label{fig:analytics_interface}
\vspace{-6mm}
\end{figure*}

\para{\textbf{Spatial Viewer}}
\phantomsection
\label{para:spatial_viewer}
The Spatial Viewer visualizes the subject locations, contexts, and the referents at the point of an action, as shown in \hyperref[fig:analytics_interface]{\NoHyper\cref{fig:analytics_interface}\endNoHyper-B}. A Trace Map displays the locations and frequency of each action for every user. It is specifically useful to infer spatial patterns of actions such as trajectory and locations of collaboration. Each action is mapped with a different color shade assigned to a user. As shown in \cref{fig:analytics_interface}, each trace point is a 3D point that represents an action overlaid with other spatial viewer components. The Spatial Viewer supports the simultaneous visualization of multiple action instances. In other words, if a range of time steps is selected in the Temporal Viewer, meaning more action instances, a more holistic spatial context of the subjects' actions can be visualized. Leveraging this technique, we visualize the full coverage of the subject's visited locations across any virtuality of XR, with only UAD-formatted recordings.

To reduce visual clutter in large collaborative multi-user scenarios, across time and space -- synchronous/asynchronous and co-located/remote, a spatial data filter is provided that allows analysts to filter users, contexts, referents, or actions. 

\begin{figure}[t]
    \centering
    \includegraphics[width=\linewidth]{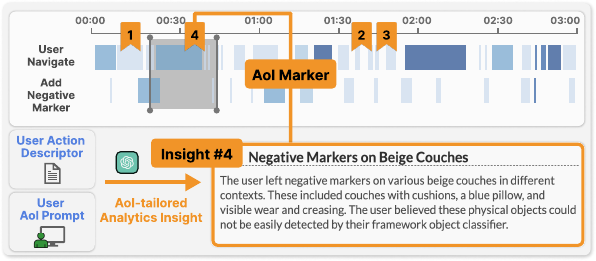}
    \vspace{-6mm}
    \caption{LLM-Analytics Assistance: Given the prompt for the direction of the analysis from the user, we generate analysis direction-tailored Analytics Insights using multimodal LLM agents. The Analysis-of-Interest Markers are associated to every insight, to pin-point the data locations of the referred insights and user's key analytics interests.}
    \vspace{-1mm}
    \label{fig:analytics_insight}
\vspace{-4mm}
\end{figure}

\para{\textbf{Temporal Viewer}}
\phantomsection
\label{para:temporal_viewer}
The Temporal Viewer (\hyperref[fig:analytics_interface]{\NoHyper\cref{fig:analytics_interface}\endNoHyper-D}) visualizes the occurrences, duration, and the frequency of actions for all subjects. Each action is a horizontal bar, where the width indicates its start and end times and color-coded using a blue sequential colormap \includegraphics[height=\fontcharht\font`\B]{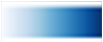} indicating the frequency of an action, where dark indicates higher action frequency. The viewer is in chronological order of the actions of the subjects. AoI Markers \includegraphics[height=\fontcharht\font`\B]{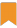} are placed above the Temporal Viewer time axis as shown in \cref{fig:analytics_insight}.
To examine actions in varying time-granularity, the analysts can also adjust the sampling interval.
To support interactivity across the analytic viewer components, selecting a time range in the Temporal Viewer updates the Spatial Viewer, Insight Viewer, and plots, aligning all information with temporal contexts. 

\para{\textbf{Data Manager}}
\phantomsection
\label{para:data_manager}
The Data Manager consists of \hyperref[fig:analytics_interface]{Data Filter}, \hyperref[fig:analytics_interface]{Data Viewer}, and \hyperref[fig:analytics_interface]{Plot Viewer}. The Data Filter consists of the list of all the actions present in the data, and provides a filter for each action. The Data Viewer presents the raw UAD information of the selected actions, including the \textit{User}, \textit{Intent}, and other information (\hyperref[fig:analytics_interface]{\NoHyper\cref{fig:analytics_interface}\endNoHyper-A}). It is useful to view the `as-is' data of actions. The \hyperref[fig:analytics_interface]{Plot Viewer} has two data visualization (\hyperref[fig:analytics_interface]{\NoHyper\cref{fig:analytics_interface}\endNoHyper-C}) that can further assist the user in comprehending the subjects' behavior information. 

\para{\textbf{Insight viewer}}
\phantomsection
\label{para:insight_viewer}
\label{para:insight_title}
\hyperref[fig:analytics_interface]{Insight Viewer} visualizes the Analytics Insights. It provides a filter for the insights that enable analysts to consolidate the insights most relevant to their interests. When a user selects an insight box, the \hyperref[para:aoi_marker]{AoI Markers} associated with that insight are highlighted. Insights are structured hierarchically for improved readability, with each insight recapitulated by a short summary (``\textit{Negative Markers on Beige Couches}'' in \cref{fig:analytics_insight}) followed by detailed content (the text below the summary headings in \cref{fig:analytics_insight}). This summary is referred to as the `Title' of an Insight.

\section{Case Study and Evaluation}
\label{sec:app_scenarios}
We design five prototype applications (A1-A5) to demonstrate the capabilities of \sys. As shown in \cref{fig:protoapps_aoi}, the applications encompass various aspects of XR, including selection, interaction, and collaboration across virtualities. We summarize the apps are explained below:


\begin{enumerate}[label=\textbf{A\arabic*}, leftmargin=5.5mm]
\vspace{-1.5mm}
    \item is a multi-user VR HMD-based application (Meta Quest Pro). The subjects are instructed to freely navigate the virtual scene and interact with the virtual objects in the scene. The application tracks their movements, gaze, and gestures.
    \vspace{-1.5mm}
    \item is an MR application using an Apple Vision Pro. The subject is given a task to select nodes of a 3D graph visualization as efficiently as possible. The selection task comprises two parts, Ray-based selection and Gaze-based selection.
    \vspace{-1.5mm}
    \item An AR application that runs on an iPad Pro. The subject is asked to scan the surroundings of their physical scene and place two types of markers. A red marker on a flat surface, and a green marker on any physical object of their preference other than the flat surface.
   \vspace{-1.5mm}
    \item We designed an AR analytics application that run on iPad Pro, for co-located subjects. The subjects are informed to openly analyze the provided data, and was encouraged to collaborate and share opinions. Both verbal and non-verbal collaboration. The application included basic tools for the analytics such as 3D barchart visualization, bar value show, and a marker for storing an interest point in the data visualization.
    \vspace{-1.5mm}
    \item This is an AR Maintenance/Inspection application that is run on an iPad Pro. The subjects of the application is instructed to anchor an AR marker on the physical locations they find interesting, and record a voice memo of the reasoning behind their placements.
    \vspace{-1mm}
\end{enumerate}

\noindent
Note that the recordings of our prototype application sessions were conducted prior to the user evaluation with a different set of user groups. Refer to \supple for further details on our prototype applications.

\subsection{User Evaluation}
\label{subsec:user_eval}
We conduct user study to evaluate \sys. Participants were asked to complete tasks using our Visual Analytics Interface, which presents the pre-recorded data of our prototype XR sessions. Below, we describe the study participants, procedure, and tasks. We term the users of the evaluation study, `\textbf{\evalsubjects}'.

\para{\textbf{Study Design}}
We recruited 14 \evalsubjects (P1-P14; 8 males, 6 females) aged 22-36 years ($\mu$=27.2, $\sigma$=3.8), from diverse fields with expertise in HCI, Visualization, XR, Computer Vision, and Systems. The \evalsubjects represent the broader user base of data analyst, referred as `Researcher' in~\cref{fig:explainable_xr_overview}. Thus, we target the study to the participants with prior data analysis experience. The study involves \evalsubjects using our Visual Analytics Interface (VAI), which displays pre-recorded data of XR sessions. For a balanced user experience assessment of VAI, we conduct the study on the \evalsubjects across varying familiarity of a visual analytics interface ($\mu$=3.1, $\sigma$=1.2; 0=unfamiliar, 5=familiar). Before the session, \evalsubjects were briefed on the functionalities of the analytics interface with a demo, and given 15 minutes to familiarize themselves. Each participant completed tasks using 5 interfaces, one for each XR prototype application, ensuring all tasks were completed by all participants. We collect their reasoning processes and usage patterns for each interface component through a task-by-task questionnaire. The session concluded with a semi-structured interview on the usability and reliability of our framework, followed by a 5-point Likert scale questionnaire for an overall assessment of \sys.

\begin{figure*}[t]
    \centering
    \includegraphics[width=\linewidth]{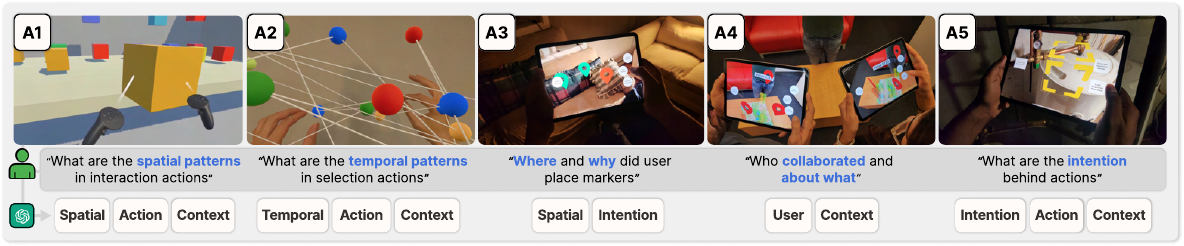}
    \vspace{-6mm}
    \caption{General-purpose Analytics Framework: \sys can be utilized for diverse analytics tasks such as pattern finding, contextual visualization, and intention grasping, of the users of an XR session. It can also provide tailored key insights based on the analyst's AoI, after assigning the appropriate LLM agent of the AoI prompt. The multifaceted insights are generated by six LLM agents specialized in Spatial, Temporal, User, Action, Context, and Intent analyses. \sys is comprehensively assessed with five prototype apps with disparate theme: (A1) VR Game, (A2) MR Selection Techniques, (A3) AR 3D Scene Reconstruction, (A4) AR Collaborative analytics, (A5) AR Maintenance/Inspection.}
    \label{fig:protoapps_aoi}
\vspace{-4.5mm}
\end{figure*}

\vspace{-0.2mm}
\para{Tasks}
Participants were given four tasks. For each task, we load the corresponding XR prototype application and configure the AoI prompt to align with the task's analysis goal (\cref{fig:protoapps_aoi}). Following are the tasks:  \textbf{T1}, \evalsubjects are instructed to use the interface to analyze action patterns of the subjects recorded in A1 and A2. \textbf{T2} focuses on analyzing the context behind the subjects' actions of our prototype app, A3. Participants are asked to describe the background scene at the time of the actions. For the third task, \textbf{T3}, \evalsubjects are asked to summarize the actions of the subjects of A4, and report the gists of subjects' collaboration if it exists. We ask about verbal and non-verbal collaboration separately. Finally, in \textbf{T4}, \evalsubjects are asked to derive the action intentions of the subjects in A5, to the best of their ability. 

\subsection{Task Results}
We present the key findings from our study that demonstrate how \sys enables comprehensive analysis of user actions and intentions in XR environments based on the data collected from the \evalsubjects.

\vspace{-0.1mm}
\para{Action Pattern Finding}
All \evalsubjects agreed (\textbf{$\mu$=4.1}) on the overall usefulness of our interface for pattern analysis (\textbf{T1}). One of the participants indicated the usability of \hyperref[fig:analytics_interface]{Insight Viewer} for pattern finding, ``\textit{The Insight viewer was incredibly helpful in judging the task performance time.}' (P4). Another participant was pleased to share their findings on the analysis of error rate for each metaphor, ``\textit{Gaze-based had more errors than the ray-based which seemed interesting to me}'' (P11). We observe that the \hyperref[fig:analytics_interface]{Plot Viewer} and the \hyperref[fig:analytics_interface]{Spatial Viewer}, were the most commonly used to complete the task. Participants were able to pin-point the exact number and duration of interactions along with the subjects' action patterns ``\textit{Subject2 interacted with 262 cubes and 27 spheres through gaze and controller, which is more than Subject1}'' (P14). 

\para{Action Context Understanding}
The \evalsubjects were overall satisfied with the the ability of the interface in visualizing the spatial context (\textbf{T2}), while displaying all the information on the subject's actions (\textbf{$\mu$=4.3}). A participant said ``\textit{I used the spatial viewer to visually get the hang of what happened and used the \hyperref[fig:analytics_interface]{Plot Viewer} to get the Referent Object name.}'' (P2), indicating the joint use of the \hyperref[fig:analytics_interface]{Spatial Viewer} and \hyperref[fig:analytics_interface]{Plot Viewer}. Another participant used the combination of \hyperref[fig:analytics_interface]{Spatial}, \hyperref[fig:analytics_interface]{Temporal}, and \hyperref[fig:analytics_interface]{Data Viewer}. ``\textit{The user placed a marker at timestamp 8/20/2024, 4:48:26PM, near the Bean bag.}'' (P1). Throughout the task, \evalsubjects relied on the class labels of physical action referents, highlighting the importance of our post-hoc \hyperref[para:action_referent_classifier]{Action Referent Classifier} spatial context reasoning. One user also emphasized the usefulness of spatial view filters: “\textit{Context, object, and tracemap to see the exact way user mapped markers}” (P10).

\para{Action Summarization}
\capitalevalsubjects correctly identified and summarized (\textbf{T3}) the collaboration between the two XR subjects, noting ``\textit{Users discussed planting trees, this was a collaborative discussion. They also discussed how wealthier people might populate places near the workplace}'' (P2). One participant, in addition to successfully completing the task, also made an insightful observation  ``\textit{ Voice memo action is for marking warnings ,if you select the whole time, and see what the user is saying, it says..}'' (P8). The study indicated that for this task, participants often selected the whole time range of a session using the \hyperref[fig:analytics_interface]{Temporal} and then use the \hyperref[fig:analytics_interface]{Data Viewer} to review the transcribed audio logs ``\textit{I selected the whole time range and then the Data Filter to focus on tasks and then the Data Viewer showed how many results there were}'' (P7). A few participants (4) relied on the \hyperref[fig:analytics_interface]{Insight Viewer} for a summary of the session ``\textit{Insight Viewer: I used it to look at the collaboration and the summarization of what each user contributed to it.}''(P12), \evalsubjects agreed on the utility of the interface for data summarized (\textbf{$\mu$=4.4}).

\para{Action Intention Inference}
The \evalsubjects actively utilized the \hyperref[fig:analytics_interface]{Insight Viewer} to infer action intentions (\textbf{T4}). ``\textit{I used insight viewer to get hints of the session. It was telling me the user tended to inspect the findings and helped figure out the user's intention.}'' (P11).  Howevers, participants often chose to not solely rely on the \hyperref[fig:analytics_interface]{Insight Viewer}, a participant opted to use the \hyperref[fig:analytics_interface]{Spatial} and \hyperref[fig:analytics_interface]{Data Viewer} to infer the intention behind the subject's action on anchoring AR Sticky note (P9). Participants seem to verify the results of the \hyperref[fig:analytics_interface]{Insight Viewer} ``\textit{I like the summaries in the insight viewer...Then I checked this through other components and the summary is correct.}''(P5). Participants agreed on the usefulness of \sys for inferring intentions behind actions (\textbf{$\mu$=4.3}).

\subsection{User Feedback}
\label{subsubsec:user_feedback}
\para{Usability}
Participants rated our Visual Analytics Interface highly for ease of use (\textbf{$\mu$=4.5}), low learning curve (\textbf{$\mu$=4.5}), and overall usefulness (\textbf{$\mu$=4.6}) across the tasks. However, one participant expressed the difficulty in using the AoI Marker ``\textit{Maybe the marker should be sorted in chronological order}'' (P13). We link each AoI Marker to one or more Analytics insights based on its relevancy to the insight. And the order is sorted based on the timestamp of the first Marker of every Insight. The \evalsubjects reported that they all jointly utilized all the components of our analytics interface, and found them useful.

\para{Reliability}
We interviewed the \evalsubjects on the output quality and usefulness of our LLM-generated Analytics Insights. All \evalsubjects found the insights helpful for data interpretation and as a foundation for building base insights (\textbf{$\mu$=4.2}) ``\textit{Yes. I used it to get an idea of the discussions between users}'' (P1), ``\textit{Overall, it is a good addition to the interface for analysis}'' (P2), ``\textit{I like the summaries in the insight viewer, it helped me learn the users' actions and interactions between users.}'' (P5). The majority did not identify any invalidity on the output of Insight viewer. One participant verified the insights using other viewers during the sessions. ``\textit{Then I checked this through other components and the summary is correct. I even didn't realize these summaries in the insight viewer were generated by AI, and I thought they were generated by human experts to help finish these tasks}'' (P5). 
However, a failure case was noted by a participant on its accuracy for analyzing temporal patterns. ``\textit{I feel some of the maths are wrong}'' (P13). Our analysis showed occasional inconsistencies in the agent's math computations, such as for the ``Average task completion time'' task. Overall, \evalsubjects evaluated the usefulness of the Analytics insights highly, and indicated a strong likelihood of using LLM-assisted analytics insights again for data analysis (\textbf{$\mu$=4.0}).
\vspace{-0.5mm}

\subsection{System Evaluation}
\label{subsec:tech_eval}
We evaluate the performance of the base setup of \sys, \hyperref[fig:log_statement]{Action Recorder, \textit{Log}}, by measuring its overhead across various XR platforms and devices. To accuractely represent the overhead, all functions were converted to synchronous calls. As \cref{tab:log_performance} indicates, the base Log function, which excludes \textit{Context} and \textit{Referent} data, has a negligible impact on XR applications \textbf{(${<}$0.14 ms)}. Even with Physical Context capture (Device camera snapshot) on an iPad Pro, application performance remains above \textbf{28.20ms (\(\sim \)35 FPS)}. For Virtual Context capture (In-application Unity camera snapshot), latency is maintained at \textbf{32.61ms (\(\sim \)30.67 FPS)}. Combining this with \textit{Referent} storage increases latency to 101.44ms due to the complexity of storing material and geometric properties of a gameobject in GLB format. However, in practical scenarios, we implement asynchronous multi-threading, reducing perceived latency to \textbf{\(\sim \)1.13ms} and minimizing the impact on XR user experience. The overhead of context snapshot capture arises from GPU readback, texture encoding, and storage. Capturing an image at 1920x1080 resolution on an iPad introduces 143.94 ms of latency. Thus, we downsample the resolution by 75\%, to 480x270, reducing the performance impact to 28.20ms (Log+PC). Downsampling is applied to both Physical and Virtual Scene snapshot capture of the \textit{Context}.

We compare the performance of a recent work, ReLive, which is comparable to \sys in functionality -- tracking user behavior data, action context (screenshots), action target referents, and supporting cross-platform environments. Under identical testing conditions (iPad Pro; averaged over 100 calls), \sys and ReLive reap similar overall performance. \sys outperforms ReLive by \textbf{\(\sim \)0.08ms} for base Log calls and achieves a negligible improvement of \textbf{\(\sim \)1.66ms} for Virtual Context capture (480x720). However, for logging with \textit{Referent}, ReLive completes the task in 1.01ms, while \sys takes 101.44ms (or 1.13ms asynchronously), due to the GLB format conversion overhead compared to the OBJ format of ReLive.



We evaluate the usefulness of LLM-generated insights within VAI and justify our choice of multi-agent approach over a single-agent approach by comparing the output quality of the two methods. The multi-agent approach decomposes the XR interaction pattern analyses into smaller sub-tasks (Spatial, Temporal, Contextual, etc.), assigns each to a specialized agent, coordinates, and ensembles to derive the best result. In contrast, the single-agent approach performs all analyses at once without task distribution. To assess the output quality, we apply the concept of self-evaluating agent~\cite{dibia2023lida, guo2023evaluating, kadavath, lin2022teaching}. We develop the evaluation metrics inspired by the SEVQ metrics of LIDA~\cite{dibia2023lida}. Our five criteria are : (\textbf{C1}) Relevance to the analysis goal, (\textbf{C2}) Compliance with the analyst's AoI prompt, (\textbf{C3}) \hyperref[para:insight_title]{Title} representation -- \textit{``How well does the \hyperref[para:insight_title]{Title} represent the insight?''}, (\textbf{C4}) Alignment with subject actions -- \textit{``How well does the insight represent the subject's actions?''}, and (\textbf{C5}) Overall diversity of insights -- \textit{``How many unique aspects are covered in an insight?''}. 

Our results (\cref{tab:llm_performance}) show that our framework generates insights well-aligned with the analyst's analytics needs (\textbf{$\mu$=9.04} out of 10; Mean of C1-C4) and provides multifaceted perspectives (\textbf{$\mu$=8.90} out of 10; C5). In all criteria, the multi-agent approach outperforms the single-agent approach, except in Criteria-3 \underline{(C3)}. Our evaluation reveals that single-agent tends to produce a more generic descriptions of insights, leading to higher score in generalizability (C3). However, \sys prioritizes constructive and specific insights, making abstract descriptions less ideal for our framework. A qualitative comparison underscores this difference: the single-agent produces outputs such as \textit{(``Inspection \underline{Log Entries}'', ``User1 left inspection logs for \underline{various} objects'')}, while multi-agent offers more detailed insights such as \textit{(``Interaction with Objects and Inspection Logs'', ``User1 interacted with objects including QR codes,..'')}. Underlined terms highlight the generic nature of single-agent outputs.
\vspace{-0.75mm}

\section{Limitation and Discussion}
\label{sec:discussion}
\vspace{-1mm}
\para{Reliance on Verbal Input for User Behavior Analysis}
EXR leverages multimodal data streams and embodied interactions to comprehensively analyze user behaviors. The UAD captures both predefined actions (e.g., Pinch, GazeAt) and \hyperref[para:action_intent_estimation]{Post Defined} actions (e.g., Speak) in a structured format. LLMs, then, contextually bridge the stream of data, and identify patterns. Most importantly, \sys can generate meaningful insights even without \hyperref[para:action_intent_estimation]{Post Defined} actions such as verbal input. For instance, in \hyperref[sec:app_scenarios]{A1}, \sys successfully analyzed user attention and spatial interactions using only gaze, gestures, and spatiotemporal data. Reliance on verbal input becomes a limitation only when the AoI of an analyst is narrowly centered on the verbal channel, such as analyzing ``\textit{Topics of discussion between users.}'' In such cases, while our visual analytics interface remains functional, LLM-assisted insights will rely solely on non-verbal data such as interacted referents, spatiotemporal patterns, and visual feeds, potentially lacking crucial contextual cues. To address this, we plan to integrate additional modalities -- physiological signals (e.g., EEG, heart rate) -- to enhance the robustness of \sys in communication-agnostic scenarios.

\para{Privacy Measures on XR Device Camera Captures}
In recent XR platforms, such as Vision OS and Meta Horizon OS, third-party apps are restricted from directly accessing device camera sensors by the OS ~\cite{appleEnterprise, jana2013enabling, kim2023erebus, kim2021design, roesner2014world}. Since we rely on camera snapshots to reconstruct 3D context point clouds, we resort to an in-application (Unity) camera snapshot to store the captures in these cases. To further address this, we plan to extend our framework to store permitted 3D scene meshes provided by the OS or middleware (e.g., ARKit).

\para{Data Recording Overhead}
We identify two sources of potential performance degradation in our Action Recorder. First, exporting a referent object involves an asynchronous GLB conversion that incurs a latency of 101.44ms. To mitigate this, we plan to perform the conversion at the end of the XR session. Alternatively, we could include an option to store referents in OBJ format, which requires less conversion overhead than GLB. While the OBJ format lacks support for PBR materials, animations, and complex hierarchical structures, the addition of an option could be beneficial for applications with performance priority. The second is due to the context snapshot capture. While we reduce the overhead by lowering the capture resolution, this degrades the quality of context point clouds. Thus, we plan to introduce asynchronous readback to preserve point cloud quality while balancing performance. 

\para{Offline Analytics Interface}
While \sys can capture and analyze diverse XR environments, it is currently limited to offline, retrospective visualization of sessions. As shown in \cref{fig:explainable_xr_overview}, our framework involves multiple processing steps including physical entity classification, context point cloud generation, and analytics insight generation, making real-time visualization challenging. In our following version of \sys, we plan to support an online visualizer that can optionally stream the session data without requiring post-processing steps.

\para{Room for Human Error in Analysis Interest Prompting}
As demonstrated through the diverse tasks of our prototype applications and evaluation, \sys can perform as a general-purpose analytics framework. Analysts can input any AoI prompt in a plain English and visualize tailored analytics insights, which help them establish base insights and preliminary hypotheses. However, we observed that our Analytics Assistant (Insight generator) fails to pin-point the useful patterns when an ambiguous AoI prompt such as ``\textit{Tell me user actions}'', is given from the analyst. It simply returned the list of referents with varying names: ``\textit{Interaction with Cube1, Cube45, Cube4}''. A prompt that is more task-specific such as ``\textit{Users’ interacted objects (properties of the objects; e.g., color, shape) across actions and their patterns}'' can output more meaningful results. We do not expect the domain researchers to prompt engineer the AoI. For our future work, we plan to integrate an agent that guides the researcher to input a structured prompt for a descriptive AoI, mitigating the possibilities of vaguely defined outputs.

\para{Incorrectness in Data Extrapolation}
We leverage LLM to deduce the intention of subjects' actions, and provide useful insights beyond the original recorded data: ``\textit{This repetitive action \underline{suggests} a detailed examination of the environment}'' (LLM output), or to infer the collaboration between users: ``\textit{This topic was a major focus, with \underline{User2 and User3 collaborating} closely on it}'' (LLM output). However, we identified that these extrapolation can suggest insights that are inaccurately derived, due to the hallucination of the LLM~\cite{duan2024llms, li2023evaluating, xu2024hallucination, yao2023llm}. As one of the solutions, we expect to enhance the correctness of the output by adopting the concept of multi-agent debate and self-correction~\cite{feldman2023trapping, huang2024towards, ji2023towards} for each agent. In addition, we plan to guide the agents to output the confidence scores of each analysis so that the researchers can assess the credibility of the insights, themselves. We also believe that this problem will be further mitigated with the advancement of reasoning ability of LLM~\cite{openaio1, openaio1Systemcard}. 


{
\renewcommand{\arraystretch}{1.2}
\setlength{\tabcolsep}{2pt}
\begin{table}[t]
\normalsize
\centering
\caption{Average overhead comparison of Log function in Action Recorder across devices and configurations. The Quest and Vision Pro do not support capture through physical device camera. 
Measurements, recorded in milliseconds, represent averages across 100 calls from 10 independent application runs. (Log: Base logging without referent or context storage; PC: Physical Context snapshot; VC: Virtual Context snapshot; R: Referent object save)}
\vspace{-2.5mm}
\label{tab:log_performance}
\begin{tabular}{| l | c | c | c | c | c |}
\hline
\textbf{Device} & \textbf{Log} & \textbf{Log+PC} & \textbf{Log+VC} & \textbf{Log+R}\\
\hline
iPad Pro & 0.08 & 28.20 & 32.61 & 101.44 \\
Meta Quest Pro & 0.13 & - & 35.04 & 82.54 \\
Apple Vision Pro & 0.14 & - & 23.64 & 72.31 \\
\hline
\end{tabular}
\end{table}
}

{
\renewcommand{\arraystretch}{1.2}
\setlength{\tabcolsep}{2pt}
\begin{table}[t]
\normalsize
\centering
\caption{Average scores per criterion for different Analytics insight generation methods, measured on a scale of 0-10. Each criterion score represents an average across 10 independent runs. (C1: Relevance to type of analysis; C2: Compliance to user's analysis-of-interest; C3: Alignment of insight to the \hyperref[para:insight_title]{Title}; C4: Alignment of insight to action; C5: Overall diversity of insights)}
\vspace{-2.5mm}
\label{tab:llm_performance}
\begin{tabular}{| l | c | c | c | c | c |}
\hline
\textbf{\rule{-2pt}{2.5ex} Method} & \textbf{\rule{-2pt}{2.5ex} C1} & \textbf{\rule{-2pt}{2.5ex} C2} & \textbf{\rule{-2pt}{2.5ex} C3} & \textbf{\rule{-2pt}{2.5ex} C4} & \textbf{\rule{-2pt}{2.5ex} C5}\\
\hline
Single-agent & 8.20 & 7.64 & \underline{9.38} & 8.72 & 8.00\\
Multi-agent \textbf{(Ours)}& 8.73 & 8.78 & 9.29 & 9.35 & 8.90\\
\hline
\end{tabular}
\vspace{-4mm}
\end{table}
}

\section{Data Privacy and Ethics}
\label{sec:ethics}
This research was conducted under IRB of the Office of Research Compliance at Stony Brook University ($1173920\_MODCR005$). All subjects provided informed consent prior to participation. To ensure privacy, data was anonymized with aliases in the \textit{User} field of the UAD, and audio recordings were transcribed to text. For our analysis, only textual interaction data was used. Captured snapshots were processed to exclude any identifiable features such as faces and name tags. No collected XR session data will be publicly released, beyond what is shared in this manuscript, to further protect subject privacy.

\section{Conclusion}
\label{sec:conclusion}
In this paper, we presented \fullsys, an end-to-end framework capable of recording, processing, and visualizing user behaviors across virtualities in diverse XR settings. Central to our approach is the User Action Descriptor, a novel format designed to integrate multimodal behavior data and action context while ensuring consistency across virtualities and multi-user scenarios. We showcased the applicability of UAD in capturing immersive sessions and showcased the usability of our LLM-assisted visual interface, which enriches the analytics experience with intelligent insights and tailored visual guidance. Through five practical XR applications, we presented a comprehensive evaluation of \sys. We envision \sys as a tool for understanding XR user experiences, enabling researchers from various disciplines to adopt XR into their studies.

\acknowledgments{We express our gratitude to Hyunji Yoon for the art work, and Matthew Castellana for video editing and narration. Also, we thank Muhammad Farrukh, Sumeer Ahmad, the anonymous reviewers, and the study participants for their valuable feedback and contributions that made this work possible. This research was supported in part by NSF award IIS2107224 and ONR award N000142312124.}



\end{document}